\documentclass[graybox]{svmult}

\usepackage{type1cm}        
\usepackage{makeidx}         
\usepackage{graphicx}        
\usepackage{multicol}        
\usepackage[bottom]{footmisc}

\usepackage{newtxtext}       %
\usepackage{newtxmath}       
\usepackage{tikz}

\usepackage{url}
\usepackage{color, colortbl}
\definecolor{Gray}{gray}{0.9}
\definecolor{Dark}{rgb}{0.69,0.76,0.87}						
\usepackage{float}
\usepackage{color}

\usepackage{hyperref}

\newsavebox\lstbox
\definecolor{gcomment}{rgb}{0.24,0.49,0.37}    
\definecolor{bleu}{rgb}{0.16,0,1}              
\definecolor{violet}{rgb}{0.49,0,0.33}         
\usepackage{listings}
\lstdefinelanguage{SoSADL}{
  morekeywords={library,via,send,receive,if,then,tell,ask,unobservable,unsaid,done,choose,or,compose,and,value,is,is abstraction,constraint,with,architecture,system,mediator,sos,datatype,function,property,gate,duty,connection,in,out,assume,behavior,guarantee,protocol,binding,unify,relay,in,suchthat,exists,implies,to,one,none,lone,some,any,all,repeat,integer,xor,forall},
  keywordstyle=\bfseries\color{violet},
  stringstyle=\color{bleu},
  morecomment=[l]{...},
  morecomment=[l]{//},
  morecomment=[s]{/*}{*/}
}
\lstdefinelanguage{DEVSNL}{
  morekeywords={passivate,start,to,in,use,when,and,receive,go,hold,after,output,has,the,range,of,'s,is,accepts,on,generates,with,type,for,time,From,perspective,sends,from},
  keywordstyle=\bfseries\color{violet},
  stringstyle=\color{bleu},
  morecomment=[l]{...},
  morecomment=[l]{//},
  morecomment=[s]{/*}{*/}
}
\lstdefinelanguage{Xtend}{
  morekeywords={def,String,IF,ENDIF,null,ELSE,FOR,ENDFOR,var,as,override,if,this,else,toFirstUpper,instanceof,instanceOf,or},
  keywordstyle=\bfseries\color{violet},
  stringstyle=\color{bleu},
  morecomment=[l]{...},
  morecomment=[l]{//},
  morecomment=[s]{/*}{*/}
}
\lstdefinelanguage{pseudo}{
  morekeywords={iterate,write,in,if,is,do,else,create,equals,for,each,to,connect,over},
  keywordstyle=\bfseries\color{violet},
  stringstyle=\color{bleu},
  morecomment=[l]{...},
  morecomment=[l]{//},
  morecomment=[s]{/*}{*/}
}

\usepackage{caption}
\usepackage{rotating}
\usepackage{lscape}

\usepackage{booktabs}

\makeindex             

\begin{document}

\title*{Teaching Simulation as a Research Method in Empirical Software Engineering
}
\author{Breno Bernard Nicolau de França, Dietmar Pfahl, Valdemar Vicente Graciano Neto, and Nauman bin Ali}
\institute{Breno Bernard Nicolau de França \at Universidade Estadual de Campinas \email{bfranca@unicamp.br}
\and Dietmar Pfahl \at University of Tartu \email{dietmar.pfahl@ut.ee}
\and Valdemar Vicente Graciano Neto \at Instituto Federal de Goiás \email{valdemarneto@ufg.br}
\and Nauman bin Ali \at Blekinge Institute of Technology \email{nauman.ali@bth.se}}
%
%
\maketitle

\abstract{
The chapter supports educators and postgraduate students in understanding the role of simulation in software engineering research based on the authors' experience. This way, it includes a background positioning simulation-based studies in software engineering research, the proposition of learning objectives for teaching simulation as a research method, and presents our experience when teaching simulation concepts and practice. For educators, it further provides learning objectives when teaching simulation, considering the current state of the art in software engineering research and the necessary guidance and recommended learning activities to achieve these objectives. For students, it drives the learning path for those interested in learning this method but had no opportunity to engage in an entire course on simulation in the context of empirical research.
}


\section{Introduction}

Several fields, such as Physics, Chemistry, Biology, and others, have matured over time and developed knowledge, theories, and laws. Emerging from this evolution, researchers and engineers created mathematical models to investigate certain conditions using computer simulation.

Software Engineering (SE), a young field of study compared to classical fields of science, has also evolved and, over time, proposed hundreds of models to simulate software development processes, software systems, and phenomena that can be observed while applying SE in practice. Such a history also yielded and consolidated methodological support for simulation in this field, including processes and guidelines, making it a relevant, contemporary research method in SE \cite{de2020role}.

Although recognized as an important research method, we still face the challenge of educating and training young researchers on when and how to apply simulation to investigate SE-related issues. Therefore, in this chapter, we discuss the teaching of simulation as a research method in SE, focusing on simulation introductory concepts and practice. For that, we frame this discussion using a simulation process as background. Moreover, we discuss several aspects related to simulation teaching in the context of SE that we believe to be important for educators and postgraduate students.  

In Section \ref{sec:simulationRole}, we start with contextualizing simulation as a research method and defining its role in SE. Section \ref{sec:learningGoals} sets up learning goals for simulation, considering mainly two contexts: i) a whole course on simulation, and ii) simulation as one of the topics in a larger course. Section \ref{sec:concepts} mentions important simulation concepts and discusses how to teach them around the notion of a process for simulation studies. Section \ref{sec:practice} discusses issues when teaching the practice of simulation and describes two simulation models used in different contexts during recent years. Section \ref{sec:lessons} presents challenges and lessons based on the authors' experience when teaching and using simulation as a research method. Additional recommended materials are described and made available in Section \ref{sec:material}, and Section \ref{sec:remarks} concludes this chapter.

\section{Simulation as a Research Method in Software Engineering}
\label{sec:simulationRole}
\begin{figure}[ht]
    \centering
    \includegraphics{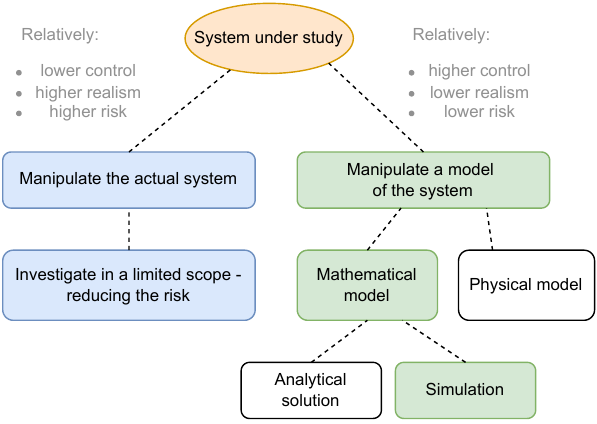}
    \caption{Role of simulation when studying a system (inspired by Law \cite{law_simulation_2013}).}
    \label{fig:SimulationsRole}
\end{figure}

Simulation, as it is understood in this chapter, is the \emph{imitation} of the behavior of a real-world process or system \emph{over time}~\cite{Banks99a}. Simulation entails developing a mathematical model mimicking a system or process's real-world behavior. We quantitatively manipulate and evaluate the mathematical model in simulation-based studies instead of changing the real-world system. Fig~\ref{fig:SimulationsRole} outlines the role of simulation. Among the different ways to manipulate and investigate a system, the main factors for a trade-off between these alternatives are realism, control, and cost/risk if the introduced change does not produce the expected results or has unforeseen consequences. In this chapter, we only consider digital models, i.e., computer programs that model the structural and behavioral aspects of the real-world phenomenon under investigation. Approaches that use physical models (e.g., scaled-down replicas of a vehicle) to manipulate a system under study for the purpose of exploration or analysis are out of scope. Similarly, mathematical models that do not have a stochastic element, where an analytical solution is more appropriate, are outside the scope of this chapter. Fig~\ref{fig:SimulationsRole} highlights the path to a simulation-based study of a system in green color.

When balancing realism, control, and cost/risk, we must weigh the trade-offs of different investigation methods. If human behaviour plays an essential role in analyzing real-world processes, no laws of nature are available on which models could be based. Thus, when realism is essential and the associated cost/risk is acceptable, real-world investigations are conducted using case studies and action research. In SE, where knowledge and evidence are strongly context-bound, realism is often mandatory, and we often have to compromise on having a lack of control in real-world settings. However, the cost and risk associated with manipulating a product in production or a development process in operation means that we want to be as confident as possible about the likely impact of a proposed change. In such situations, simulation emerges as an economical and proactive tool to forecast outcomes before committing resources to change. Suppose the simulation results do not contradict the expectations based on made assumptions. In that case, there is a higher plausibility that the envisioned changes might also work in usually much more complex real-world settings \cite{Pfahl2014}. To what degree the behaviour of the simulation model matches the real-world behavior of the simulated process or product should be stated as a requirement upfront to manage expectations. If a simulation model must closely mimic real-world behavior, thorough validation is essential. This may include expert judgment, empirically derived reference modes to which the simulation output is compared, and other types of checks \cite{law_simulation_2013}.

Simulation is an invaluable tool for supporting knowledge acquisition and decision-making in research and practice. It enables what-if analyses, even in cases with limited data or a lack of established theories. Simulations can assist by modeling assumptions and approximations, offering objective judgment on proposed solutions. However, it is essential to use caution to avoid over-interpreting results.

In the context of SE, simulation mostly refers to two different types of systems under study: the software (or system) development process, i.e., \textit{process simulation} \cite{Pfahl2014}, or the software product (or system), i.e., \textit{product simulation}. Recently, the term \textit{digital twin} has gained popularity \cite{Eramo2022}. While digital twins are essentially simulation models, they distinguish themselves by representing an existing physical artifact (product) or real-life process. Moreover, a digital twin is constantly fed with data collected from, often in real-time, and used for behaviour monitoring and prediction. In this book chapter, we do not restrict ourselves to digital twins but consider the full spectrum of simulation models, including those that are built to analyse and explore the behavior of products and processes that do not yet exist - or, at least that deviate from existing ones and are created to assess the effects of implementing planned changes.

Process simulation refers to the simulation of software development processes capturing the flow of work products from requirements to the end product and explaining the use of resources (people and tools) along the process \cite{Müller2008}. The goal is usually the exploration of improvement opportunities or the evaluation of the effects of a proposed improvement (i.e., change) of the process organisation overall or one or more process elements. More broadly, Kellner et al. identified six categories of purposes of process simulation, i.e., strategic management, planning, control and operational management, process improvement and technology adoption, understanding, and, last but not least, training and learning \cite{KELLNER1999}. The application domains range from requirements engineering to testing and investigating the effect of specific process changes in a single development phase to analyzing multi-project contexts and software maintenance and evolution \cite{Zhang2010}. A process simulation model would typically capture properties of activities, artifacts, and resources involved in the process. In the context of SE, a process simulation model could, for example, explore the effects of changing the number and/or skills of developers allocated to specific tasks (e.g., coding and testing) on the quality of the outputs of these tasks (e.g., production code and test code) \cite{Pfahl2014}.

Product simulation refers to the simulation of a software product or a system, including embedded systems. The goal is usually to evaluate the behaviour of the software product or system, for example, in the context of testing. With the rising popularity and fast evolution of AI-based systems, simulation might also play a role in training artificial neural networks (ANNs). Assume, for example, that a cyber-physical system, such as an autonomous vehicle controlled by a trained ANN, is modeled as a digital twin that moves in a simulated virtual world. Then, in the same way, the physical vehicle's ANN might be trained using the data collected by sensors in the real world, and the digital twin's ANN might be trained using the data collected by simulated sensors in the simulated virtual world. If the digital twin, as well as the virtual world, closely resemble the physical vehicle and physical world, it should not make a difference whether the ANN is trained using real-world data or data from the simulated virtual world. Both aspects, training of ANNs as well as simulation-based safety testing, have become hot topics in the context of automated driving systems (ADS). In recent surveys of approaches for testing ADS \cite{Khan2023, Tang2023}, simulation-based testing has been identified as an essential method to speed up the testing process by facilitating the exploration of high-risk driving scenarios and the automatic execution of many different test scenarios. 

In combination with powerful and often domain-specific simulators, such as CARLA\footnote{https://carla.org} in the specific context of ADS development and testing, both process and product simulation have not only become important in the industry but more and more are seen as research tools, e.g., for exploring the effectiveness of principles of communication and collaboration among software developers, for exploring how design principles may support the maintainability and evolution of software, or for exploring which test scenario selection principles make the testing of ADS most effective and efficient. The potential of simulation as a research tool naturally makes simulation relevant for learning.

\section{Setting up Learning Objectives for Simulation}
\label{sec:learningGoals}

Teaching simulation as a research method in SE is challenging, as several undergraduate programs overlook simulation and modeling introduction courses. In the ACM Curriculum Guidelines for undergraduate degree programs in Computer Science, computer simulation is part of a broader discipline called Computation Sciences \cite{10.1145/2534860}. It aims to enable competencies such as formal modeling of real-world situations and use such models in simulations at an introductory level \cite{cc2020}. More advanced courses in this reference curriculum are considered elective, encompassing different purposes of simulation, tradeoffs among model attributes such as accuracy and complexity, specific modeling and simulation techniques, verification and validation (V\&V), and supporting tools. 

This scenario sets a general context for teaching computer simulation as an SE research method in graduate courses. These courses should consider basic and advanced simulation concepts and particular techniques and approaches often used in SE.

As background, the students need previous knowledge of (i) programming and algorithms, as the notions of sequential instructions for reaching a goal is also important in simulation, as well as the notion of abstractions, variables, and control flow; (ii) introduction to statistics, mainly, concepts regarding data distributions and equation modeling; in terms of research methods, (iii) an overview of ESE, experiments, and threats to validity is also important. This last pre-requisite differentiates teaching simulation as a technique and an empirical research method. Furthermore, it makes such a course more suitable for postgraduate students. In the case of focusing on the technique perspective, simulation can be (and is referred to by ACM) an undergraduate course.

In line with the ACM Curriculum Guidelines, we provide the learning objectives for a full course on Simulation, considering its use as a research method in Empirical Software Engineering (ESE). As a terminal learning goal, the student should be able \emph{to recognize situations in which simulation is an appropriate research method and apply it accordingly, supported by existing processes and guidelines, when adopting well-established simulation approaches}. Specific or enabling learning objectives include:

\begin{enumerate}
    \item Explain the relationship between modeling and simulation, i.e., thinking of simulation as dynamic modeling. [Understand]
    \item Differentiate among the different types of simulations, including \emph{in virtuo} and \emph{in silico} environments. [Understand]
    \item Explain the benefits of simulation and modeling in SE research. [Understand]
    \item Choose an appropriate modeling approach for a SE problem or situation. [Evaluate]
    \item Create a new or extend an existing model of a SE phenomenon and use that model in a simulation. [Apply] 
    \item Verify and validate the results of a simulation, raise and address potential validity threats. [Analysis]
    \item Perform simulation experiments comparing different scenarios and explain any differences. [Apply]
\end{enumerate}

Another usual scenario is to include simulation as a topic in other courses, e.g., in an ESE course. In such contexts, covering all the topics at their proposed learning levels will be unfeasible. So, the level should be adjusted for each object to accommodate the course expectations. For instance, Course 4 (see Section \ref{sec:experiences}) establishes a level of `Understand' with the main simulation-related concepts to understand the whole simulation life cycle (see Section \ref{sec:concepts}) and approaches for software engineering (System Dynamics and DEVS). Additionally, two `Evaluate' objectives are useful for i) decisions when using simulation as a research method and ii) to assess the quality of this sort of study based on guidelines \cite{de2016experimentation} and the Empirical Standards \cite{ralph2021empirical} on Quantitative Simulation. This way, the students can review and assess primary studies involving simulation for their theses, but they do not learn how to create models using any particular simulation approach as it would demand several classes for the 'Apply' objectives. Regarding the 'Apply' level, the students in Course 4 receive training on performing experiments using an existing model (usually a Systems Dynamics model).


\section{Experiences on Teaching Simulation}
\label{sec:experiences}

This section describes the main experiences of teaching simulation as a research method in SE. Sections \ref{sec:concepts} and \ref{sec:practice} mention these experiences from a set of different courses organized by the authors, including: \textbf{Course 1}: a course on simulation for graduate program students (PhD candidates and master's students) (2019); \textbf{Course 2:} a tutorial taught in the European Conference on Software Architecture (2021)\footnote{Slides and supplementary material: https://ww2.inf.ufg.br/insight/tutorialecsa2021/} \cite{NetoECSA} and in the Brazilian Congress on Software (2021); \textbf{Course 3:} a summer school on the topic (2023); and, \textbf{Course 4:} an ESE course including simulation as a unit.

\textbf{Course 1:} This 60-hour presential elective course was taught to 13 master's and Ph.D. candidates during the first academic semester in 2019 (March to July 2019). The dynamics were based on (i) expositive classes (16 hours) about DEVS associated with practical programming activities in the laboratory, (ii) student seminars (20 hours), and (iii) practical project development, scientific paper writing, and presentation (24 hours). The learning objectives were to (i) provide subsidies for students to develop studies and projects based on systems simulation, (ii) introduce the fundamentals of modeling dynamic systems through the use of simulations, (iii) present notations, formalisms, and tools used for this purpose and (iv) enable the students to specify simulation models for representing systems in different domains. The program was based on teaching DEVS with MS4Me, some seminars on chosen articles, and other simulation technologies and paradigms, such as NetLogo for ABS. Besides executing partial practical programming exercises over the first half of the course, during the second half of the course, while they were presenting seminars, they were also required to develop a practical project that should be presented at the end of the course and also communicated in a scientific paper to be submitted to a workshop. Five scientific papers were developed in groups and accepted in the workshop that year.

textbf{Course 2:} This is a short course of three hours in the format of a tutorial taught twice: at the European Conference on Software Architecture (April 2021) and at the Brazilian Congress on Software (September 2021). The audience was students of several levels, professors, researchers, and practitioners who attended those events. Those tutorials were remote due to the COVID-19 pandemic. The content was based on an expositive presentation about the importance of simulation in the German industry by Dr. Pablo Antonino (1 hour) and expositive lessons on DEVS in the last two hours. Material and instructions on downloading MS4Me and sample examples were made available to the attendees through a website. The learning objectives were \cite{ecsaNeto21} to teach (i)  knowledge of fundamentals of smart-ecosystem architectures, which was the background of the course (ii) Introduction of DEVS for newcomers using MS4Me, (iii) Introduction of a simulation approach (Dynamic-SoS \cite{Manzano20}), to show how we could automatically generate DEVS code with architectural reconfiguration support; and (iv) Practical experience of smart-ecosystem simulation, guiding the attendees on how to deploy, run and interpret the results of the simulation.

\textbf{Course 3:} This was a 1-week remote intensive course of 16 hours during summer in Brazil (February 2023\footnote{\url{https://summerschool.inf.ufg.br/en/index.html}}). The audience was students and researchers on many levels. More than 40 people from the entire country attended the course. We had an 8-hour course on DEVS with MS4Me, besides a Tutorial with Guidelines for In-Silico Experiments with Simulation (4 hours). In the end, we had a workshop on scientific papers running in parallel with an entire-day hackathon financially sponsored by RTSync\footnote{\url{https://rtsync.com/}} to award the best projects and also technically sponsored by TriboCloud\footnote{\url{https://www.tribocloud.com/}}, who provided a cloud environment for running the simulator.

\textbf{Course 4:} This is a one-semester ESE course for graduate (PhD and Master) students. It has been offered for seven years. Its main learning goals are (i) to recognize the established empirical methods, (ii) to properly select research methods, and (iii) how to conduct primary (planning) and secondary (whole process) studies in practice. The course is organized into units that correspond to research methods. One of them is \textit{In Silico} studies, mostly simulation. Before this unit, the students are introduced to ESE, including its origins, paradigms, threats to validity, and other related concepts; controlled experiments; and quantitative (statistical) analysis, which includes measurement scales, descriptive statistics, exploratory data analysis, and statistical inference. This unity includes definitions and motivations for \textit{in silico} studies; simulation-based studies life cycle and their main activities; particularities of simulation for SE; exemplary studies and models in discrete-event and system dynamics; and guidelines for simulation in SE. Expositive classes, discussions, and seminars are used for activities. The first assessment is the elaboration of a research protocol for a primary study, in which the students define a research goal and should be able to select the appropriate methods to conduct the research. Additionally, one last assessment is the execution of a secondary study. In these assessments, simulation can be used as a research method (primary study), or simulation studies must be assessed in the quality appraisal (secondary studies).

Finally, we are also aware of other initiatives, for instance, teaching ARENA\footnote{\url{https://www.rockwellautomation.com/pt-br/products/software/arena-simulation.html}} as part of a regular SE course in a Brazilian university \cite{pinheiro2022}.

\section{Teaching Simulation Concepts}
\label{sec:concepts}

To set up our perspective on simulation, we adopt the following definition from ~\cite{Banks99a}:
``Simulation is the \emph{imitation} of the operation of a real-world process or system \emph{over time}. Simulation involves the generation of an \emph{artificial} history of the system, and the observation of that artificial history to draw \emph{inferences} concerning the operating characteristics of the real system that is represented''.

Besides the definition, it is important to present and discuss the situations in which simulation suits well for SE research (Section \ref{sec:simulationRole}). Also, presenting motivations for adopting simulation by considering its advantages, such as the low (financial, safety, cost, or even ethical) risk of performing experiments in a highly-controlled environment, anticipating ``what-if'' scenarios, knowledge gaining and theory-building, and possibly working with failure scenarios. Conversely, it is equally important to highlight disadvantages or challenges, such as possible lack of knowledge to develop a simulation model, lack of data for calibration, developing toy or simple models that are not accurate enough for the established research goals, or unfeasibility of validating complex/ambitious models \cite{de2020role}.  

\begin{figure} [htbp]
    \begin{center}
        \includegraphics[width=0.9\textwidth,height=0.4\textheight,keepaspectratio]{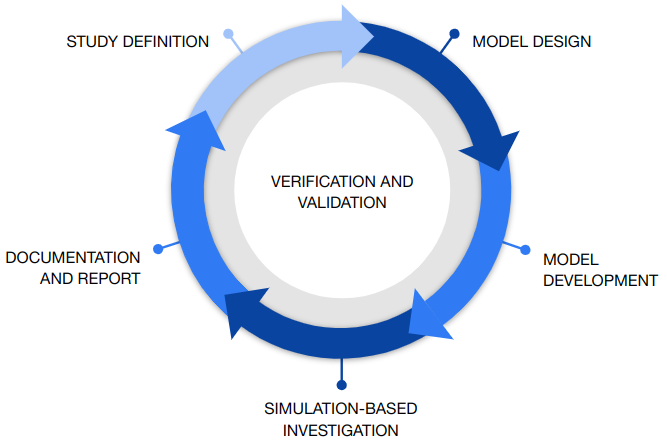}
    \end{center}
    \caption{Simulation-Based Studies Life Cycle (adapted from \cite{de2020role}).}
    \label{fig:SimLC}
\end{figure}

Simulation-based studies need methodological support, such as guidelines and processes \cite{ali2012consolidated}, which are important for planning, designing, execution, analysis, and reporting. de Fran{\c{c}}a and Ali \cite{de2020role} have discussed the role of such studies in SE research. We present their proposed process, which they further complement with supporting guidelines in Fig~\ref{fig:SimLC}. 
We use this process as a framework to structure our discussion of several simulation concepts below. 

The first step in the process, the study definition, is similar to other empirical investigations. It includes setting research goals, formulating research questions, and describing the context. 

For the next step of model design and development, the students must be exposed to concrete approaches, such as discrete-event simulation (e.g., DEVS) and system dynamics, which are the most common in SE. 

Many different simulation approaches exist, with their characteristics, strengths, and weaknesses \cite{NetoKassab2023}. The most appropriate approach depends on the simulation goal and the aspects of the system to be simulated, as they capture different characteristics to be represented or highlighted in the simulation model. Some of the most common approaches (not being restricted to) include:

\begin{enumerate}
    \item \textbf{Discrete Event Simulation (DEVS):}  DEVS simulation models capture the structure and behavior of a system as a series of events that occur over discrete units of time. It is often used to model complex systems, such as manufacturing plants or transportation systems, where the order of events is important.
    \item \textbf{Continuous Simulation:} This simulation approach replicates the continuous evolution of a system's behavior and finds frequent application in modeling physical systems, particularly those involving fluids and thermodynamics, where the system's state undergoes constant alterations.
    \item \textbf{System Dynamics (SD):} This is a particular continuous modeling technique focusing on capturing and analyzing the behavior of dynamic systems by incorporating the fundamental principles of feedback and non-linearity. This approach plays a pivotal role in understanding and predicting the dynamics of complex systems characterized by a significant degree of uncertainty. SD finds widespread application in various domains, primarily industrial, management, finance, and social sciences.
    \item  \textbf{Agent-Based Simulation (ABS):} A simulation approach designed to emulate the intricate dynamics of systems composed of autonomous agents, which can encompass individuals, businesses, or even animals. ABS has found extensive application across various domains, notably economics, social sciences, and engineering.
\end{enumerate}

Considering the associated concepts, the dynamics of these approaches can be taught in traditional expositive classes and readings (see Section \ref{sec:material}), but the teacher can also use educational games. For example, to learn system dynamics, there is the \emph{Beer Game}\footnote{\url{https://en.wikipedia.org/wiki/Beer_distribution_game}}, invented by Jay Forrester, available in multiple formats and is quite simple to apply. Additionally, starting modeling examples and exercises from simpler models and increasing complexity is an important strategy to not overwhelm students with the complexity of large models and non-trivial problems or phenomena to be simulated. Larger and more complex models may be explored once the students have captured the basic concepts of the intended approaches and can be used to work on projects for more practical aspects (see Section \ref{sec:practice}). Finally, including models representing software processes and products is important for the students to realize the range of possible simulated phenomena in SE.

Verification and validation (V\&V) of dynamic simulation models include different procedures, most independent of the simulation approach since they take models as a `black box'. Theoretically, it is important to differentiate verification from validation \cite{babuska2004verification}. The first is associated with checking implementation consistency and correctness against its specification or conceptual model. Instead, the second implies checking if simulation outcomes resemble real-world scenarios regarding accuracy according to simulation goals.

Conceptually, it is important to introduce V\&V procedures, how to apply them, and understand when to use each procedure, including their prerequisites in model assumptions, operationalization, and available data. Additionally, each V\&V procedure evaluates a model considering particular aspects, and, in this sense, they are complementary, and using more than one procedure is desirable. These procedures are crucial in mitigating threats to validity since the simulation model is the main source of threats in this type of study. Not all procedures have established guidelines; some provide only the main idea. This way, presenting examples from the literature on how these procedures were applied to different models is interesting for providing analytical capabilities for the students.

Simulation-based investigations are about performing experiments using valid simulation models. So, it requires some knowledge of experimental design and quantitative analysis (Statistics). If students do not have such background, it is important to provide an overview of experimental terminology and relevant designs (e.g., factorial designs and sensitivity analysis). As experimental designs for simulation can be defined in an \emph{ad-hoc} way or based on existing metamodels \cite{de2016experimentation}, the students need to understand the consequences of missing factors (or model inputs) or alternatives (possible values for factors), i.e., using incomplete designs. This tradeoff can be explored in classes using a simple model and exemplifying candidate designs for experimenting with that model.

Documentation and reporting become simple if the simulation-based study is properly planned. There are guidelines for planning and reporting simulations \cite{de2016experimentation}. 

Teaching all topics and concepts in each process phase can involve expositive classes, readings, discussions, seminars, and theoretical exercises. One interesting exercise, which can be proposed individually or in groups to foster discussion, is to analyze published studies against the process and the guidelines. For operationalizing it, after an expositive class introducing the concepts, the instructor previously selects good and poor examples or simulation studies, assigns them to students, provides the process and guidelines descriptions, and asks for a presentation in seminars where the students should describe the study in an overview along with a critical analysis based on the provided guidelines. This activity allows familiarization with simulation concepts, analysis using a concrete case (study), and critical thinking. 

A lesson from giving classes about simulation in SE for years is that exposing students only to concepts in an abstract way, without showing concrete studies, model examples, and potential threats to validity \cite{de2015simulation}, will not unleash their full potential and the critical thinking required for future researchers or experimenters.

\section{Teaching Simulation Practice}
\label{sec:practice}

Teaching simulation practice needs a reference system or behavior so that students can work with modeling and simulation approaches. In this sense, having datasets with reference behaviors or system descriptions may support working with small projects and exercises. Alternatively, instead of developing simulation models from scratch, students may 
start from existing models and then evolve these models to answer new questions.

One needs a simulation approach, language, and tools apart from the real system to develop a simulation model. Section \ref{sec:concepts} mentions some simulation approaches, but we remark that several of those may cohabit with the same simulation tool and can be combined to deliver more elaborated results, such as a combination of continuous and discrete simulations (that can be connected through co-simulation or even hybrid simulation models). For instance, MS4Me is a proprietary simulator for DEVS based on Eclipse, but it has a free academic license for students. In MS4Me, we can combine Monte Carlo and DEVS in the same simulation model. Similarly, AnyLogic\footnote{\url{https://www.anylogic.com/}} supports multiple approach models, combining constructs from discrete-event, system dynamics, and agent-based simulation in a composite model.
Moreover, we also highlight that some simulation approaches are more appropriate than others when supporting or matching the characteristics of the problem being solved. For instance, SD can be more suitable than DEVS when we intend to model causal relationships due to its inherent feedback loops explicitly described in a model, in contrast to hidden dynamics in event-discrete models, which focus on detailing measurable aspects of entities \cite{sweetser1999comparison} \cite{tako2009comparing}. On the other hand, if one needs to solve a problem in which there is a composition of entities and the system behaviors are triggered by events, DEVS (as the tool for the solution space) is an abstraction closer to the problem space, reducing the need to a more complex mapping between problem and solution spaces. 


The following subsections (\ref{sec:devs} and \ref{sec:dsbrooks}) present two examples of models and how they have been used to support teaching simulation studies.

\subsection{A Discrete-Event Model of a Flood Monitoring System}
\label{sec:devs}

When teaching DEVS analogously to other computational techniques, we need to progress from simple instances until we reach more complex scenarios. DEVS is a formalism based on atomic and coupled models \cite{Zeigler:2012}. Atomic models can represent individual entities, such as a software component or an entire system. In contrast, coupled models are combinations of atomic (and other coupled) models, such as an entire system or multiple systems interoperating.

Atomic models comprise the following elements \cite{hicss2018}: (i) ports (input and output); (ii) a labeled state machine that executes transitions in response to input and output events (hence governing the system operation); (iii) functions that can be used to process data; (iv) data types; and (v) events. Coupled models are expressed as a System Entity Structure (SES), a formal structure governed by axioms that express how atomic models communicate with each other \cite{Zeigler:2012}. 

\begin{figure}[!ht]
 \centering
  \includegraphics[scale=0.55]{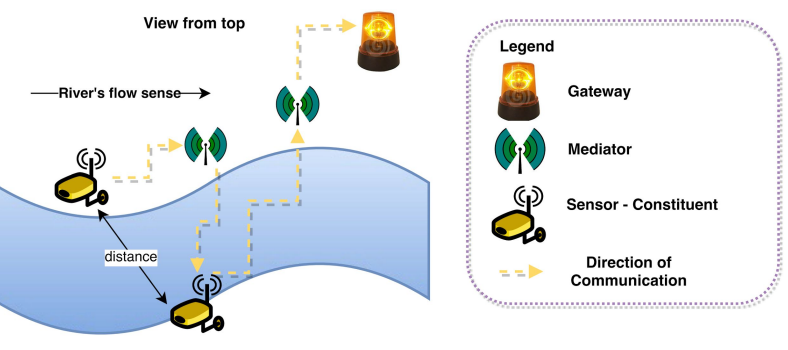}
  \caption{A Flood Monitoring System as part of a Smart City System \cite{WDES2016}.}
     \label{FMSoSArchSimul}
      \end{figure}

In a prior post-graduation (master's and Ph.D. students) course, the motivational example used to help was a Flood Monitoring System (FMS)\footnote{The Flood Monitoring System is a recurring practical scenario that has already been covered in other prior works. In this work, we highlight that we used the scenario as inspiration and modeled the system directly in DEVS. Also, this work differs from previous works that used ASAS \cite{hicss2018}, which is not a formalism but rather a method for automatically generating DEVS code from SoSADL, that is, a method that involves two formalisms. For the chapter purpose, we reuse the same motivational scenario and describe the corresponding DEVS codes, but not the SoSADL codes.}, a system intended to be part of a smart city. FMS monitors rivers crossing urban areas, which pose a great danger in rainy seasons, potentially damaging property, threatening lives, and spreading disease. FMS notifies possible emergencies to residents, business owners, pedestrians, and drivers located near the flooding area, as well as governmental entities and emergency systems. The FMS we describe concerns a single behavior: \textit{flood alert}.  Sensors are spread on the river's edges with a regular distance among them, and mediators exist between every pair of sensors at a pre-established distance between them. The data collected by sensors are collected and transmitted until they reach the gateway. In a flood, the gateway emits an alarm for public authorities. FMS comprises two sensors, two mediators, and one gateway, as in Figure~\ref{FMSoSArchSimul}.

Listing \ref{devsnlExemplo} illustrates the code in DEVS of a mediator using the DEVS Natural Language dialect. Data types are defined in Lines 1-22 in Listing \ref{devsnlExemplo}. Duties (which designate gates of mediators) with their respective connections are defined on Lines 8-16. The individual behavior of this mediator is specified in lines 18-23. In short, this behavior defines that after receiving the coordinates of constituents that it mediates (lines 19-20), this mediator will repeatedly receive data from one sensor (line 22) and forward them to the other (line 23). Similarly to gates, duties are transformed into inputs of ports in the atomic model (lines 24-27 in Listing \ref{devsnlExemplo}). Finally, the behavior is translated into automata in the atomic model (lines 29-42 in Listing \ref{devsnlExemplo}).

\begin{lstlisting}[caption={An atomic model for a Mediator generated in DEVS.},label=devsnlExemplo, linewidth={9cm},numbers=left, stepnumber=1, firstnumber=1, numberfirstline=true,language=DEVSNL]
A Distance has a value!
the range of Distance's value is Integer! 
use distance with type Distance!

A Abscissa has a value!
the range of Abscissa's value is Integer! 
use abscissa with type Abscissa!
A Ordinate has a value!
the range of Ordinate's value is Integer! 
use ordinate with type Ordinate!
Coordinate has x and y!
the range of Coordinate's x is Abscissa!
the range of Coordinate's y is Ordinate!
use coordinate with type Coordinate!

A Depth has a value!
the range of Depth's value is Integer! 
use depth with type Depth!
Measure has coordinate and depth!
the range of Measure's coordinate is Coordinate!
the range of Measure's depth is Depth!
use measure with type Measure!

accepts input on FromCoordinate with type Coordinate!    		
accepts input on ToCoordinate with type Coordinate!    		
accepts input on FromSensors with type Measure!    		
generates output on Measure with type Measure!    		

to start hold in s0 for time 1!
hold in s0 for time 1!
from s0 go to s1! //Unobservable
passivate in s1!
when in s1 and receive Coordinate go to s2!    		
passivate in s2!
when in s2 and receive Coordinate go to s3!    		
passivate in s3!
when in s3 and receive Measure go to s4!    		
hold in s4 for time 1!
after s4 output Measure!
from s4 go to s5!		
hold in s5 for time 1!
from s5 go to s3! //Unobservable
\end{lstlisting} 

In DEVS courses, the first challenge the students usually face is synchronizing the output and input events in both the sender and receiver when modeling two systems or modules exchanging messages. Since both are often represented using atomic models, they are governed by state machines whose state transitions can occur due to input and output events. Since they also occur in specific time frames settled by the modeler, a message exchange between the atomic models is only well-succeeded when, in the same instant in time, the sender is sending a message and the receiver is in a state that, when receives some data, triggers the state transition and some processing. Otherwise, it does not work, and the models are not paired accordingly. To facilitate that, we developed a type of pattern for programming in DEVS so that the students could reduce their errors at that phase \cite{SAC2018}.

Another important difficulty perceived during the first moments with DEVS is related to the specification of labeled state machines. DEVS has a formal foundation; some students or practitioners are not necessarily used to this type of notation. Then, examples that help with that difficulty are welcome. 

When the students overcome this first moment, in general, the newcomers shall understand other details related to the programming resources of the simulation tool, such as receiving multiple messages from different senders at the same moment, and other advanced concepts in DEVS, such as the specification of ports, the manipulation of the content in messages being exchanged between the atomic models, the events processings, selective running, and other features that can be specific to the tool adopted, such as Aspects and Specialization. After these steps, the student is prepared to run a simulation specification process and use the conceived models for experimentation and benchmarking, besides other glimpsed potential applications.

\subsection{A System Dynamics Model of the Brooks Law}
\label{sec:dsbrooks}

Another example of a simple and intuitive model to use in classes with a different simulation approach is the model proposed in Chernoguz \cite{chernoguz2011system}. It captures the essence of Brook's Law, which states that ``\emph{adding manpower to a late software project makes it later}'' \cite{brooks1995mythical}.

\begin{figure}[!ht]
 \centering
  \includegraphics[scale=0.3]{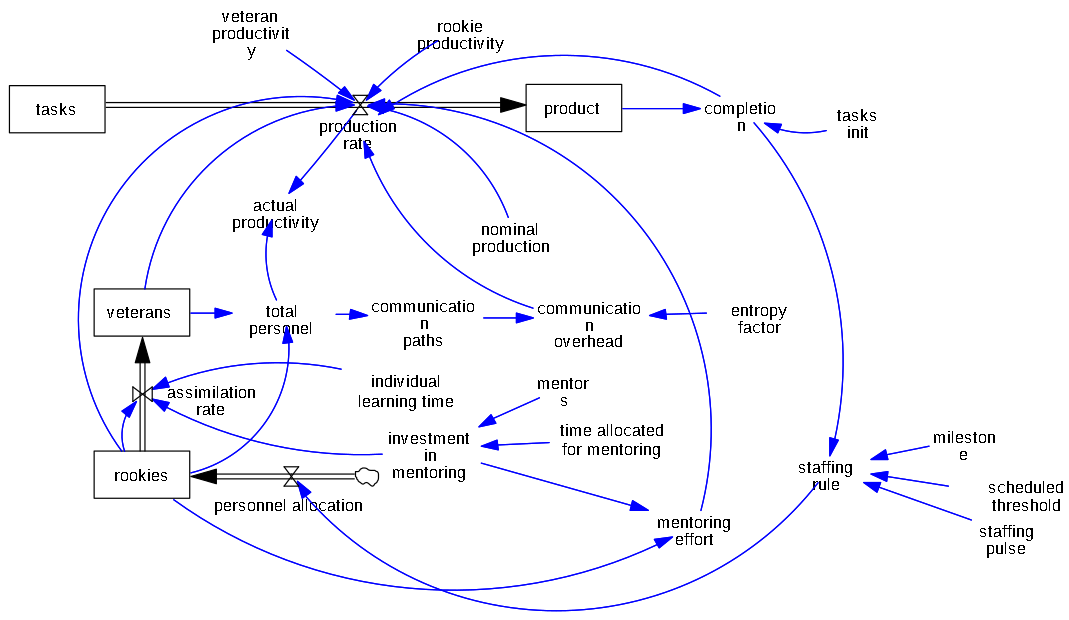}
  \caption{An System Dynamics Model of the Brook's Law (based on \cite{chernoguz2011system}).}
\label{fig:SDBrooks}
\end{figure}

Figure~\ref{fig:SDBrooks} depicts the SD diagram of the model, including the \emph{stocks}, i.e., variables in squares that can accumulate values, and \emph{flows}, which are variables looking like a valve linked to stocks so that their values can be increased or decreases based on a rate. The model is implemented using the Vensim PLE tool. Both stocks and flows can be defined in terms of constants or functions. For instance, the stock \emph{rookies} represents the new additional staff allocated using the flow \emph{personnel allocation} rate. This amount of rookies in the stock variable evolves to the stock \emph{veterans} as these gain experience with learning and mentoring through the flow \emph{assimilation rate}. In runtime, the SD engine uses differential equations over those definitions to determine the values at each constant and regular simulation step. All the others are called supporting or auxiliary variables, for example, the variables \emph{individual learning time} and \emph{investment in mentoring}. 
 
This sort of model and diagram is useful for explaining the phenomena under simulation. The causal loops among variables form a chain of reasoning (a form of causal mechanism), so the input changes can be explained via model traces to the output. Of course, the tool abstracts the complexity of the system dynamics execution and formalisms for the students so they can focus on the model variables and their relationships.  

In this model, variables that do not depend on other variables are understood as independent. This way, they can be used as input parameters of the model or, experimentally, as model factors. For example, \emph{staffing pulse} is the variable responsible for defining the number of new members to be added at a particular simulation time, based on a \emph{scheduled threshold} that defines the amount of delay to trigger the allocation of additional manpower.

In its original form, this SD model allows the performance of several experiments to understand the effects of allocation strategies, team size, level of team experience, and other variables on development productivity. Additionally, the students may explore model enrichment with additional variables, representing more detailed parts of this phenomenon and even coupling this model to a larger software process/project model. 

An example of a simulation-based investigation follows the experimental design in Table \ref{tab:expdesign}. In this design, we consider two factors (supporting variables): staffing pulse and entropy factor at four and two levels, respectively. This would lead us to a factorial design with eight runs/trials, considering this is a deterministic model (with no stochastic components). The values for staffing pulse levels were exploratory, considering a baseline (0), and for the entropy factor, we adopted the nominal values from \cite{chernoguz2011system}. 

\begin{table*}
\caption{Example of Experimental Design}
\label{tab:expdesign}
\begin{tabular}{@{}l@{}l@{}l@{}}
\toprule
\multicolumn{1}{c}{\textbf{Trial}} & \textbf{Staffing Pulse} & \multicolumn{1}{c}{\textbf{Entropy Factor}} \\
\midrule
\multicolumn{1}{c}{1} & \multicolumn{1}{c}{0} & \multicolumn{1}{c}{0.03} \\
\multicolumn{1}{c}{2} & \multicolumn{1}{c}{2} & \multicolumn{1}{c}{0.03} \\
\multicolumn{1}{c}{3} & \multicolumn{1}{c}{4} & \multicolumn{1}{c}{0.03} \\
\multicolumn{1}{c}{4} & \multicolumn{1}{c}{6} & \multicolumn{1}{c}{0.03} \\
\multicolumn{1}{c}{5} & \multicolumn{1}{c}{0} & \multicolumn{1}{c}{0.06} \\
\multicolumn{1}{c}{6} & \multicolumn{1}{c}{2} & \multicolumn{1}{c}{0.06} \\
\multicolumn{1}{c}{7} & \multicolumn{1}{c}{4} & \multicolumn{1}{c}{0.06} \\
\multicolumn{1}{c}{8} & \multicolumn{1}{c}{6} & \multicolumn{1}{c}{0.06} \\
\bottomrule
\end{tabular}
\centering
\end{table*}

Defining and evolving an experimental design with a concrete model is also useful for the students to understand the consequences of selecting particular types of designs, such as the complexity of the output analysis and the number of runs. This reinforces concepts about controlled experiments and allows the students to experiment with different designs and realize their advantages and disadvantages in practice.

\begin{figure}[!ht]
 \centering
  \includegraphics[scale=0.4]{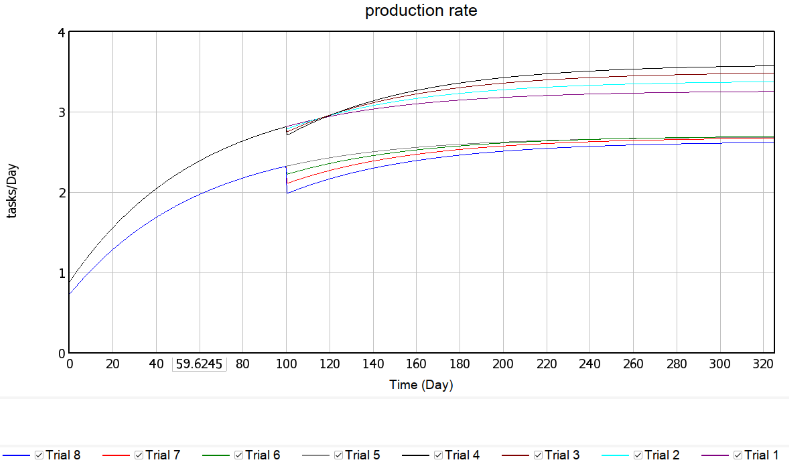}
  \caption{Output of the production rate variable for eight trials.}
\label{fig:8runsplot}
\end{figure}

Figure~\ref{fig:8runsplot} presents the output of the eight scenarios defined in the design. At all the trials, except for the baselines with no pulse, the behavior is similar: as soon as the project is detected as behind the schedule (at time 100), the staff increases according to the pulse (2, 4, or 6 new members) and it generates an immediate decrease in the production rate. As the rookies gain experience, the production rate increases again until it overcomes the former production rate. This is in line with the statement of Brooks himself when he refers to the law as an ``outrageous oversimplification'', as the law is applicable in certain conditions, i.e., it is observed mostly in the short term. 

The point of this example is not to show a perfect model but to show how students can gain valuable insights from simulating phenomena of interest, gaining knowledge from a ``trial and error'' or exploratory approach, both in the modeling and experimentation perspectives. 

\section{Challenges and Lessons Learned}
\label{sec:lessons}

The challenges and lessons provided herein come from different experiences from the authors when teaching simulation, including perspectives in which simulation was taught as (i) an empirical/experimental research method or (ii) a general-purpose tool to solve problems, particularly related to critical domains.  
\\\\
\noindent \textbf{Define your learning goals for simulation based on your course goals.} Consider your audience and the course goals when setting up the learning goals for Simulation. Our experience ranges from simulation as a topic in an ESE course to a full course on Simulation as a general-purpose tool. These settings will differ regarding hours and how in-depth the concepts and practice can go for the students or using larger projects, for example. 
\\\\
\noindent \textbf{Run simulations in class!} Even if your learning goals do not include practical knowledge of modeling or experimenting with models, it is important to show concrete models to students and run simulations during the classes so they can realize the concepts with a simple but solid example. As models selected for teaching are relatively small or simple, educators and students may have no big issues when running these models.
\\\\
\noindent \textbf{Teach how to move from the problem to the solution space.} During an introductory course on simulation, it can be really important to teach how to map concepts from the problem space to the solution space. In other words, abstraction exercises are key for modeling and simulation. For instance, in DEVS, all the problem space concepts must be mapped to atomic models, coupled models, ports, events, and behaviors expressed in a labeled state machine (in the solution space). Hence, several difficulties can emerge, such as mapping the granularities of the problem space concepts to the solution space concepts and converting an intended system behavior to an event-driven state machine or causal loop formalism since many students may lack such knowledge.
\\\\
\noindent \textbf{Suitability of simulation notations for SE practitioners and students.} A survey study with 58 software engineers (in which more than half had not had any contact with simulation models before) revealed that simulation formalism could not be that intuitive for them \cite{LebtagTSVN21}. The practitioners envisioned several opportunities for improvement, including reinforcements in the visual representation, user support, and model animation, adding support for colors, and the possibility of adding their symbols to represent the explored domain better. The professionals argued that such improvements could reinforce the scalability of the models (visual presentation and execution) and the management of the models and increase the models' reuse for use in fast prototyping practices. Although visual elements are relevant in this case, semantics play an important role in making it suitable or not for a specific audience. The more the notation exposes its mathematical abstractions, the more complex and error-prone it becomes for users. In general, simulation notations were developed for systems engineers. Hence, the notation or user experience often does not conform to what software engineers use. Transferring this challenge to young researchers (students) in SE is fair. As a consequence, this fact leads to the next lesson learned.
\\\\
\noindent \textbf{A process to incorporate simulation into practical projects.} While teaching simulation, particularly to software engineers, it is important to teach a basic process that they can follow to incorporate simulation models in their projects \cite{Franca2021}. We are aware of SE teams from Brazil and Germany that have adopted the following process \cite{NetoECSA}: Step 1 - Understanding the problem, Step 2 - A mapping between problem and space solution, Step 3 - A system specification, Step 4 - Design of the simulation model by mapping the system specification into the simulation formalism basic concepts, Step 5 - Simulation execution, observation, and logging, Step 6 - Analysis of results, and Step 7 - System development itself. The adoption of such a process was positive for both teams. They use it in industrial project development in partnership with large companies and the Fraunhofer IESE Institute in Germany. In Brazil, it was used to generate successful simulations in projects in the Smart Cities and Space domains.
\\\\
\noindent \textbf{Highlight the importance of having domain experts available for students.} As shown in Figure~\ref{fig:SimLC}, verification and validation are ongoing throughout the lifecycle of a simulation-based study. Validation is often done by reproducing the reference behavior and obtaining feedback from the domain experts. For example, domain expert involvement is necessary for deciding and confirming the scope of the investigation, i.e., deciding which questions the model should be able to answer, operationalizing the model with data from an organization, and documenting and communicating the results. The expert's involvement throughout the life cycle also ensures we achieve a certain buy-in and trust in the model. Which is essential for a simulation-based investigation.

The specification of simulation models of the Brazilian space system in DEVS was performed at the Brazilian National Institute for Space Research (INPE). The expert in Space systems provided his/her domain expertise, which allows us to model the system precisely \cite{SAC2018}. Moreover, the expert also acted as an evaluator for the quality and reliability of the simulation model. 
\\\\
\noindent \textbf{V\&V is key for experimentation, but modeling is essential for gaining knowledge.} There is little to no value in trying to provide evidence from a simulation model that has not been validated yet. Knowing the main source of validity threats in a simulation-based study is the model itself; students must be aware that one single V\&V procedure is not enough to validate a model. On the other hand, the modeling process in which students will incrementally gain knowledge about the process, system, or phenomenon to be simulated is valuable even initially with a simple simulation model; students can have insights and even discard unfeasible ideas or unrealistic assumptions.
\\\\
\noindent \textbf{Simulation-based investigations are experiments and should be systematically designed.} Although some simulation outcomes may look reasonable, and changing model parameters and running simulations seem simple and easy, students must know that experimenting with models is not about ``fishing'' the combinations that look better for a given research question. Exceptional scenarios, factor interactions, and sensitivity analysis should be considered a systematically designed experiment. This way, students need to practice designing experiments based on model parameters.
\\\\

\section{Recommended Readings and Teaching Material}
\label{sec:material}

\subsection{Readings}

\begin{itemize}
    \item A comprehensive chapter on the role of simulation in software engineering research \cite{de2020role}.
    \item The Modeling and Simulation Body of Knowledge (MSBoK) is also a recent asset and an important reading \cite{oren2023body}.
    \item A set of planning and reporting guidelines for simulation-based studies \cite{de2016experimentation}.
\end{itemize}
 
\subsection{Slides}

All the slides are available at the Zenodo repository: \url{https://zenodo.org/records/11544898}
\begin{itemize}
    \item Slides of a complete introductory course on DEVS with MS4Me.
    \item Slides of an introductory on modeling and simulation for software architectures.
    \item Slides of an introductory view of \emph{in silico} studies, including simulation-based ones.
\end{itemize}

\subsection{Working Simulation Models}

Several simulation models are also available at the Zenodo repository: \url{https://zenodo.org/records/11544898}
\begin{itemize}
    \item Two elementary functional DEVS models (a Hello World) and a reduced version of the Flood Monitoring System as discussed in Section \ref{sec:devs}.
    \item Some DEVS simulation models related to prior publications.
    \item The simulation model described in Section \ref{sec:dsbrooks}.
    \item The Vensim website also publishes some sample models, available at: \url{ https://www.vensim.com/documentation/sample_models.html}
\end{itemize}

\subsection{Supporting Tools}

\begin{itemize}
    \item DEVS is one of the most traditional formalisms in M\&S \cite{Zeigler2023} and has different free and proprietary implementations. MS4Me is a proprietary Java/Eclipse-based application that RTSync offers, with a free one-year license and lifelong license (in case you purchase it) and some licenses for academic purposes.
    \item There is also an implementation in Python \cite{van2014modular}, and many others that can be consulted\footnote{\url{https://en.wikipedia.org/wiki/List_of_discrete_event_simulation_software}}.
    \item Vensim is a simulation environment for System Dynamic models. It has the PLE version for educational purposes and other paid versions.
    \item AnyLogic is a multi-approach, proprietary environment for simulation, including system dynamics, discrete-event, and agent-based simulation.
\end{itemize}

\subsection{Additional Teaching Material}

\begin{itemize}
    \item A reporting standard and evaluation checklist\footnote{\href{https://github.com/acmsigsoft/EmpiricalStandards/blob/master/docs/QuantitativeSimulation.md}{Simulation checklist}} for assessing the quality of a simulation-based study in SE \cite{ralph2021empirical}.
    \item Some material is available at: https://ww2.inf.ufg.br/insight/tutorialecsa2021/
    \item The Beer Game to learn System Dynamics is available at \url{https://beergame.org/the-game/}. There are instructions for a board or table game. Also, they have a software version of the game for Windows. 
\end{itemize}

\section{Final Remarks}
\label{sec:remarks}

This chapter discusses how teaching simulation could support students, practitioners, and researchers in using it as a research instrument. For that purpose, we compiled a set of important definitions for background, discussed what we understand as the main teaching simulation concepts and principles, and communicated some recommendations based on our prior experience while teaching simulation. We also discussed what we believe to be the main challenges and lessons learned, besides providing supplementary material that teachers can use to teach simulation. We hope newcomers and teachers will consume this material to learn and teach simulation in software engineering.

\begin{acknowledgement}
This work has been supported by ELLIIT, a Strategic Area within IT and Mobile Communications, funded by the Swedish Government. The result has also been supported by a research grant for the GIST project (reference number 20220235) by the Knowledge Foundation in Sweden.
\end{acknowledgement}
%

\bibliographystyle{plain}
\bibliography{references}
\end{document}